\documentclass{frontiersSCNS} 
\usepackage{url,hyperref,lineno,microtype,subcaption}
\usepackage[onehalfspacing]{setspace}
\DeclareRobustCommand{\ion}[2]{\textup{#1\,\textsc{\lowercase{#2}}}}

\def\keyFont{\fontsize{8}{11}\helveticabold }
\def\firstAuthorLast{Wedemeyer {et~al.}} 
\def\Authors{Sven Wedemeyer\,$^{1*}$, Gregory Fleishman\,$^{2}$, Jaime de la Cruz Rodr\'iguez$^3$, Stanislav Gun\'ar$^{4}$, Jo\~ao M. da Silva Santos$^{5}$, Patrick Antolin$^{6}$, Juan Camilo Guevara G\'omez\,$^{1}$, Mikolaj Szydlarski\,$^{1}$, Henrik Eklund\,$^{1,3}$}

\def\Address{$^{1}$ Rosseland Centre for Solar Physics, Institute of Theoretical Astrophysics, University of Oslo, Postboks 1029 Blindern, N-0315 Oslo, Norway\\
$^2$ Center For Solar-Terrestrial Research, New Jersey Institute of Technology, Newark, NJ 07102, USA\\
$^{3}$ Institute for Solar Physics, Dept. of Astronomy, Stockholm University, AlbaNova University Centre, SE-10691 Stockholm, Sweden\\
$^{4}$ Astronomical Institute, The Czech Academy of Sciences, 25165 Ond\v rejov, Czech Republic\\
$^{5}$ National Solar Observatory, 3665 Discovery Drive, Boulder, CO, USA\\
$^{6}$ Department of Mathematics, Physics and Electrical Engineering, Northumbria University, Newcastle upon Tyne, NE1 8ST, UK }

\def\corrAuthor{{Sven Wedemeyer}}

\def\corrEmail{sven.wedemeyer@astro.uio.no}

\def\apj{ApJ}
\def\apjl{ApJL}
\def\apjs{ApJS}
\def\aap{A\&A}
\def\aaps{A\&AS}
\def\solphys{Sol. Phys.}

\def\mnras{MNRAS}

\def\pasj{PASJ}
\def\ssr{Space Sci. Rev.}
\def\araa{Annual Review of Astronomy and Astrophysics}

\def\araa{ARA\&A}
\def\apj{ApJ}
\def\apjl{ApJ}
\def\apjs{ApJS}
\def\apss{Ap\&SS}
\def\aap{A\&A}
\def\aapr{A\&A~Rev.}
\def\aaps{A\&AS}
\def\mnras{MNRAS}
\def\pasj{PASJ}
\def\solphys{Sol.~Phys.}
\def\ssr{Space~Sci.~Rev.}

\begin{document}
\onecolumn
\firstpage{1}
 
\title[Numerical modelling of the Sun at millimetre wavelengths]{Prospects and challenges of numerical modelling of the Sun at millimetre wavelengths}

\author[\firstAuthorLast ]{\Authors} 
\address{} 
\correspondance{} 
\extraAuth{}
\maketitle 

\begin{abstract} 
The Atacama Large Millimeter/submillimeter Array (ALMA) offers new diagnostic possibilities that complement other commonly used diagnostics for the study of our Sun. In particular, ALMA's ability to serve as an essentially linear thermometer of the chromospheric gas at unprecedented spatial resolution at millimetre wavelengths and future polarisation measurements have great diagnostic potential. 
Solar ALMA observations are therefore expected to contribute significantly to answering long-standing questions about the structure, dynamics and energy balance of the outer layers of the solar atmosphere. In this regard, current and future ALMA data are also important for constraining and further developing  numerical models of the solar atmosphere, which in turn are often vital for the interpretation of observations. The latter is particularly important given the Sun's highly intermittent and dynamic nature that involves a plethora of processes occurring over extended ranges in spatial and temporal scales. 
Realistic forward modelling of the Sun therefore requires time-dependent three-dimensional radiation magnetohydrodynamics that account for non-equilibrium effects and, typically as a separate step, detailed radiative transfer calculations, resulting in synthetic observables that can be compared to observations.  
Such artificial observations sometimes also account for instrumental and seeing effects, which, in addition to aiding the interpretation of observations, provide instructive tools for designing and optimising ALMA's solar observing modes. 
In the other direction, ALMA data in combination with other simultaneous observations enables the reconstruction of the solar atmospheric structure via data inversion techniques.
This article highlights central aspects of the impact of ALMA for numerical modelling for the Sun, their potential and challenges, together with selected examples. 

\tiny
 \keyFont{ \section{Keywords:} Sun: radio radiation, Sun: atmosphere,  Sun: magnetic fields, Radiative transfer} 
\end{abstract}

\section{Introduction} 
\label{sec:intro}

When pointed at the Sun, the Atacama Large Millimeter/submillimeter Array \citep[ALMA,][]{2009IEEEP..97.1463W} mostly observes radiation that originates from the solar chromosphere. This atmospheric layer, which is situated between the photosphere below and the transition region and corona above, is highly dynamic and intermittent and shows variations on a large range of spatial and temporal scales. 
Plasma with chromospheric conditions can also be found in the corona in the form of prominences and coronal rain. These structures are integral components of the solar corona in the sense that they do not only reflect specific physical processes of the corona but also influence its evolution \citep{Vial_Engvold_2015ASSL..415.....V, Antolin_Froment_10.3389/fspas.2022.820116}. The investigation of the thermodynamic conditions and morphology of these dense and cool structures supported by the magnetic field is therefore also a major field in which ALMA can make a major contribution.

Despite very active research regarding the chromosphere, which involves observations at many different wavelength ranges supported by numerical simulations, yet many fundamental questions concerning this layer remain open. The main reason is that the chromosphere is notoriously difficult to observe. Only a small number of spectral lines and continua are formed in the chromosphere, usually across extended height ranges. The formation of chromospheric spectral lines involves non-equilibrium effects such as, e.g., non-local thermodynamic equilibrium \citep[NLTE\footnote{In this context, NLTE or non-LTE  describes deviations from LTE conditions for the atomic level populations (which can then be calculated under the assumption of statistical equilibrium), the electron density (mostly due to  non-equilibrium hydrogen ionisation),  and the the radiative source function that is no longer given by the Boltzmann function.}, see, e.g.,][and references therein]{1955psmb.book.....U,1978stat.book.....M,1992ApJ...397L..59C} and time-dependent hydrogen ionisation \citep{2002ApJ...572..626C}. Consequently, the few currently available diagnostics like the spectral lines of singly ionised calcium and magnesium are difficult to interpret, in particular in combination with instrumental limitations. As a result, the physical properties of the observed atmospheric region can only be derived with rather large uncertainties, hampering the progress in understanding this important part of the solar atmosphere.

Observations of the solar continuum radiation with ALMA, as offered on a regular basis since 2016, provide unprecedented diagnostic possibilities that are complementary to other chromospheric diagnostics 
\citep{2002AN....323..271B,2011SoPh..268..165K,2012IAUSS...6E.205B,2016SSRv..200....1W,2018Msngr.171...25B}.
The radiation continuum at sub-millimetre/millimetre ((sub-)mm) wavelengths, including the range accessed by ALMA, forms essentially under conditions of local thermodynamic equilibrium (LTE) so that the observed brightness temperature, $T_\mathrm{b}$, is closely related to the actual (electron) temperature of the chromospheric gas in a (corrugated) layer whose average height roughly increases with the selected observing wavelength. 
Unfortunately, observations at (sub-)mm wavelengths prior to ALMA had too low spatial and temporal resolution for resolving the small spatial and short temporal scales on which the intricate chromospheric dynamics occur. 
For instance, the Berkeley-Illinois-Maryland Array (BIMA) had a spatial resolution corresponding to a restored beam size of $\sim 10$'' at a wavelength of $\lambda = 3.5$\,mm \citep[][]{2006A&A...456..697W,2009A&A...497..273L}. 
The full diagnostic potential of millimetre wavelengths has therefore only been unlocked by ALMA thanks to its high temporal and (comparatively high) spatial resolution, which significantly exceeds the resolution achieved by previous millimetre and radio observatories. 
It should be noted that the millimetre wavelengths addressed in this article are also referred to in terms of their corresponding frequencies (a few 10s\,GHz up to $\sim$1\,THz) and as (ALMA) receiver bands\footnote{\url{https://www.eso.org/public/teles-instr/alma/receiver-bands/}} that covered the discussed wavelength range. In particular, ALMA Bands~3 and 6, which have been used most frequently for solar observations so far, refer to (central) wavelengths of 3.0\,mm and 1.3\,mm and corresponding frequencies of 100\,GHz and 230-240\,GHz, respectively. 

Because ALMA is relatively new as a diagnostic tool for the solar chromosphere, still many aspects are not understood well yet. For instance, the exact formation heights and thus the layers sampled by the different receiver bands of ALMA and likewise the oscillatory behaviour seen in the ALMA observations are still debated 
\citep[see, e.g.,][]{2021RSPTA.37900174J,2020A&A...634A..86P,2022A&A...661A..95N,2021A&A...652A..92N}.
On the other hand, as solar observing with ALMA is still in its infancy, its capabilities will continue to improve in the near future. However, any new ability to obtain unprecedented observations in any part of the spectrum always brings its own challenges in the interpretation of the resulting data. Fortunately, diagnostics and understanding of the ALMA observations and their relationship to the coordinated observations in other spectral domains can benefit from dedicated numerical modelling. 

Like in many other fields of astrophysics, numerical simulations have developed into an essential tool in solar physics. Also in the context of solar observations with ALMA, simulations help to interpret observational data but can also be used to develop and optimise new observing strategies \citep[see, e.g.,][]{2007A&A...471..977W,2015A&A...575A..15L,2021ApJ...914...52F}. In return, comparison with observations provide crucial tests for the veracity of existing models of the solar atmosphere.  
In this brief overview article, the potential value of numerical simulations for solar science with ALMA is explored, ranging from forward modelling of thermal and non-thermal mm continuum radiation and the impact of magnetic fields (Sect.~\ref{sec:forwardmodelling}) to data inversion techniques and modelling of instrumental and seeing effects  (Sect.~\ref{sec:processing_analysis}). Examples of scientific applications are presented in Sect.~\ref{sec:examples}, followed by a summary and outlook in Sect.~\ref{sec:summary}. 

\section{Forward Modelling and Artificial Observations} 
\label{sec:forwardmodelling}

Forward modelling of the solar atmosphere is typically split into the following steps: (i)~Radiation (magneto)hydrodynamics simulations (Sect.~\ref{sec:rmhd}), (ii)~synthesis of observables via radiative transfer calculations (Sect.~\ref{sec:rt}), and (iii, optionally) application of simulated observational effects (e.g., limited angular resolution, see Sect.~\ref{sec:degrad}).

\begin{figure*}[t]
	\centering
	\includegraphics[width=0.92\textwidth, trim=0 0.1cm 0 0,clip]{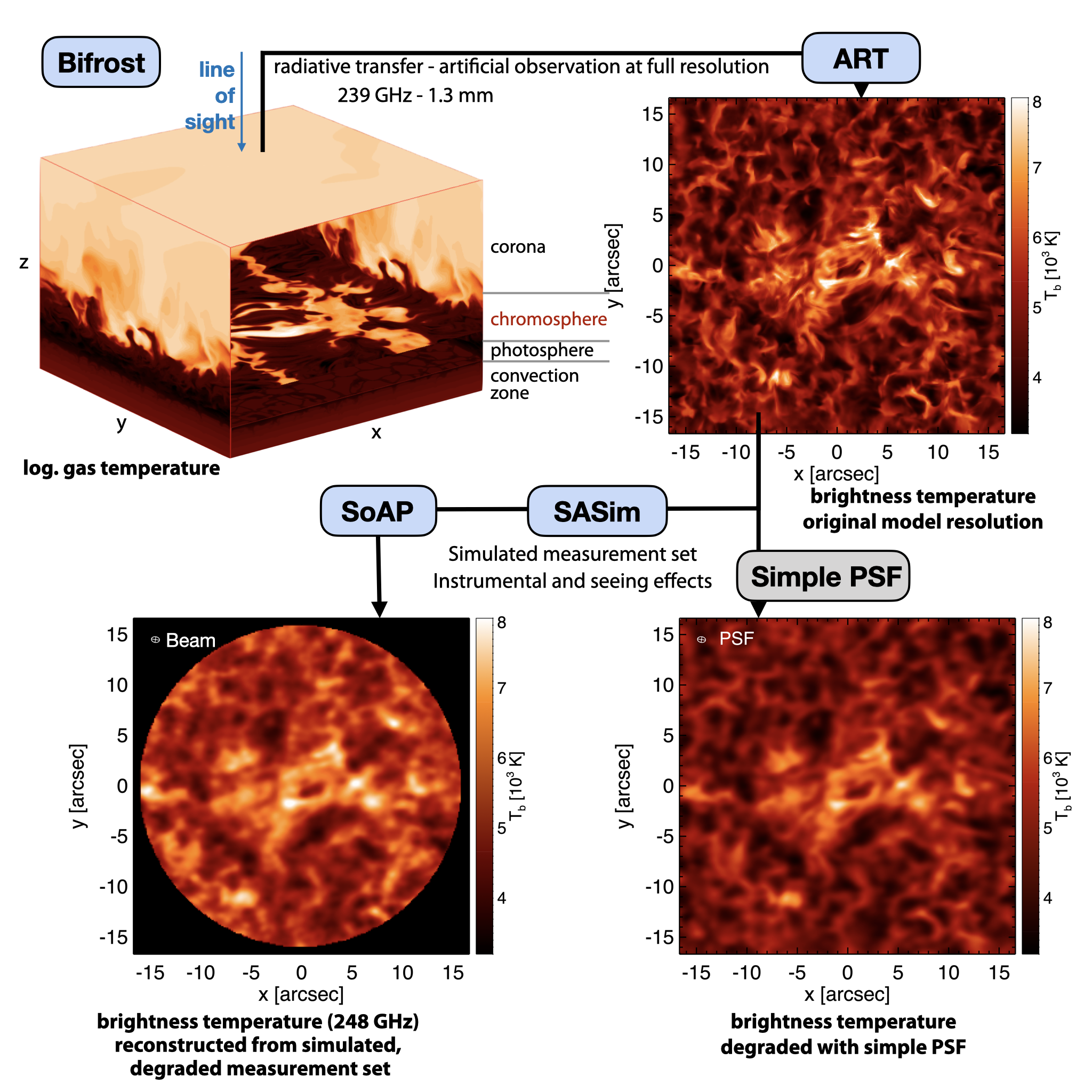} 
    \caption{Forward modelling of synthetic brightness temperatures at millimetre wavelengths starting from a 3D model snapshot (or a time series of such) from a Bifrost simulation (top left). The model features an enhanced network region in the centre surrounded by Quiet Sun. The radiative transfer calculations with ART produce a brightness temperature map for the selected frequencies (shown here: 239\,GHz, top right) to which then a simple point spread function (PSF)  equivalent to an idealised synthesised beam for ALMA can be applied (lower right). The PSF is shown in the top left corner of the panel. A more realistic simulation of atmospheric seeing and instrumental effects as done with SASim (and then reconstructed with SoAP) results in a stronger degradation (lower left) of the original image than compared to the simple PSF. Please note that only a moderate example is shown that corresponds to good observing conditions. The synthesised beam is shown in the top left corner of the reconstructed image.}
	\label{fig:bifrost-art-sasim}
\end{figure*}

\subsection{Radiation magnetohydrodynamic simulations} 
\label{sec:rmhd}

Semi-empirical models of the solar atmosphere like those by \citet{1981ApJS...45..635V}, \citet{1993ApJ...406..319F}, \citet{2008ApJS..175..229A} and others have been an important milestone and still are widely used for reference. 
Millimetre continuum observations were also used for the construction of these models, which therefore  give a first idea of where and under which conditions the radiation continuum at different millimetre wavelenghts is formed -- \emph{on average}. 
While the employed radiative transfer modelling, which even accounts for NLTE, is elaborate, this class of models can by nature not account for the pronounced temporal and spatial variations seen at the much increased resolution of modern observations. 

The next step in the development towards realistic models was therefore to account for temporal variations in the chromosphere. The time-dependent one-dimensional simulations by \citet{1995ApJ...440L..29C}
and variations therefore capture well the dynamics introduced by shock waves that propagate through the solar atmosphere and the resulting implications for the chromospheric plasma properties, i.e. the ionisation degree of hydrogen and thus the (non-equilibrium) electron density \citep{2002ApJ...572..626C}. These simulations have been used for the synthesis of the millimetre continuum and thus provided first predictions of the brightness temperature variations that a telescope with sufficient resolution would be able to observe \citep[][]{2008Ap&SS.313..197L}. 

However, the solar atmosphere and in particular the highly dynamic and intermittent chromosphere is a truly time-dependent three-dimensional phenomenon, which poses significant challenges for realistic modelling capable of reproducing observational findings. Also, temporal variations on short timescales are typically connected to spatial variations across short length scales. Consequently, accounting for the full time-dependence and multi-dimensionality of the solar chromosphere is a substantial step forward from one-dimensional approaches. In view of limited computational resources, early 3D simulations were restricted in the overall number of grid cells, seeking a compromise between required resolution and extent of the computational domain, and the physical processes that could be numerically treated \citep[see, e.g.,][]{2000ApJ...541..468S,2004A&A...414.1121W}. The enormous increase in computational power over the last decades now enable simulations with much higher numbers of grid cells and thus a better representation of the chromospheric small-scale structure and larger extents of the modelled region. However, self-consistent numerical simulations of whole Active Regions are still at the modelling frontier \citep{2009ApJ...691..640R}. 

Numerical two-dimensional (2D) and and three-dimensional (3D) models produced with the radiation magnetohydrodynamics (rMHD) simulation codes Bifrost \citep{2011A&A...531A.154G} and \mbox{CO$^5$BOLD}  \citep{2012JCoPh.231..919F} have already been used as basis for the synthesis of mm continuum radiation 
\citep[see, e.g.,][]{2007A&A...471..977W,2015A&A...575A..15L} but an increasing number of codes is developing the necessary functionality, e.g., MURaM \citep{2022arXiv220403126P}. 
Both Bifrost and \mbox{CO$^5$BOLD}  solve the equations of magnetohydrodynamics and radiative energy transfer together with a realistic equation of state and realistic opacities and further relevant physics. A typical model includes a small part of the solar atmosphere \citep[from a few Mm to a few 10\,Mm, cf.][and references therein]{2016SSRv..200....1W} and extends from the upper convection zone into the chromosphere and/or low corona (see Fig.~\ref{fig:bifrost-art-sasim}, upper left). 
This way the dynamics in the model are driven self-consistently and all layers mapped by ALMA are included. A simulation typically starts with an evolved model snapshot (or any other initial condition) and is evolved in time step by step, where the computational time steps are of the order of 1\,ms to 100\,ms, depending on the magnetic field strength in the model. 
Simulation snapshots of the physical parameters can be output at freely selectable cadence. 
Modelling the layers of the solar atmosphere above the temperature minimum in a realistic way requires the inclusion of additional physical processes and deviations from equilibrium conditions that are usually computationally expensive. As discussed in Sect.~\ref{sec:rt_thermal}, the detailed treatment of time-dependent non-equilibrium hydrogen ionisation, like it is implemented in Bifrost, is of particular importance for the continuum radiation at millimetre wavelengths. 
Adding also non-equilibrium ionisation of helium and ion-neutral interactions (ambipolar diffusion)
significantly increases the computational costs.  
Consequently,  only a small number of models so far can account for these additional ingredients and are necessarily limited to 2.5D in order to render such modelling computationally feasible \citep{2020ApJ...891L...8M}.   
These models  suggest that the effective formation heights of the millimetre continuum in both ALMA Band~3 and 6 is similar in active regions (ARs) and network regions, which contradicts results from previous simulations \citep[see][and references therein]{2016SSRv..200....1W} and actual ALMA observations \citep[e.g.,][]{2022arXiv220508760H}. 
Clearly, the inclusion of more physical processes relevant under chromospheric conditions as implemented in the 2.5D simulations by \citet{2020ApJ...891L...8M} is an essential step in the right direction. However, given the complicated small-scale dynamics of the chromosphere, modelling this layer in \emph{full 3D at sufficient resolution} is a  critical requirement that comes with high computational costs. 
 
\subsection{Spectrum synthesis}
\label{sec:rt}

In the Sun, the continuum radiation is mostly due to thermal emission mechanisms as discussed in Sect.~\ref{sec:rt_thermal} but non-thermal emission needs to be taken into account  in flares (see Sect.~\ref{sec:rt_nonthermal}). As discussed below, this results in particular requirements for the forward modelling of meaningful millimetre radiation.  

\subsubsection{Thermal radiation}
\label{sec:rt_thermal}

At (sub-)mm wavelengths, the main source of opacity is due to thermal free-free absorption (bremsstrahlung) from electron-ion free-free encounters and H$^-$ \citep[see, e.g.,][]{1985ARA&A..23..169D,2006A&A...456..697W,2016SSRv..200....1W}. 
In other words, the free-free emission originates when free thermal electrons collide with ambient ions and atoms, the total amount of which depend on the electron and ion density, chemical composition, and temperature. 
All these three inputs vary strongly in the solar atmosphere.
In particular, the thermal radiation continuum is sensitively dependent on the local electron density (or line-of-sight electron density).  
In the presence of strong magnetic field, additional contribution from gyro emission \citep{2019ApJ...880L..29A} needs to be taken into account (see below). 

Many radiative transfer codes commonly use electron densities derived from local plasma properties such as the gas temperature to calculate the thermal free-free emission under the assumption of instantaneous LTE conditions. 
As demonstrated convincingly by \citet{2002ApJ...572..626C}, the rapid variations of the plasma properties in the solar chromosphere due to propagating shock waves in combination with finite hydrogen ionisation and recombination time scales result in significant deviations of the electron density from equilibrium values. A realistic calculation of synthetic continuum intensity maps at millimetre wavelengths therefore requires detailed non-equilibrium electron densities as input \citep[see, e.g.,][]{2006A&A...460..301L}.  However, including non-equilibrium hydrogen ionisation in numerical 3D rMHD simulations is computationally expensive, resulting in only a small number of adequate simulations so far \citep{2011A&A...531A.154G,2015A&A...575A..15L,2016A&A...585A...4C}.

\begin{figure*}[t]
	\centering
	\includegraphics[width=0.91\textwidth]{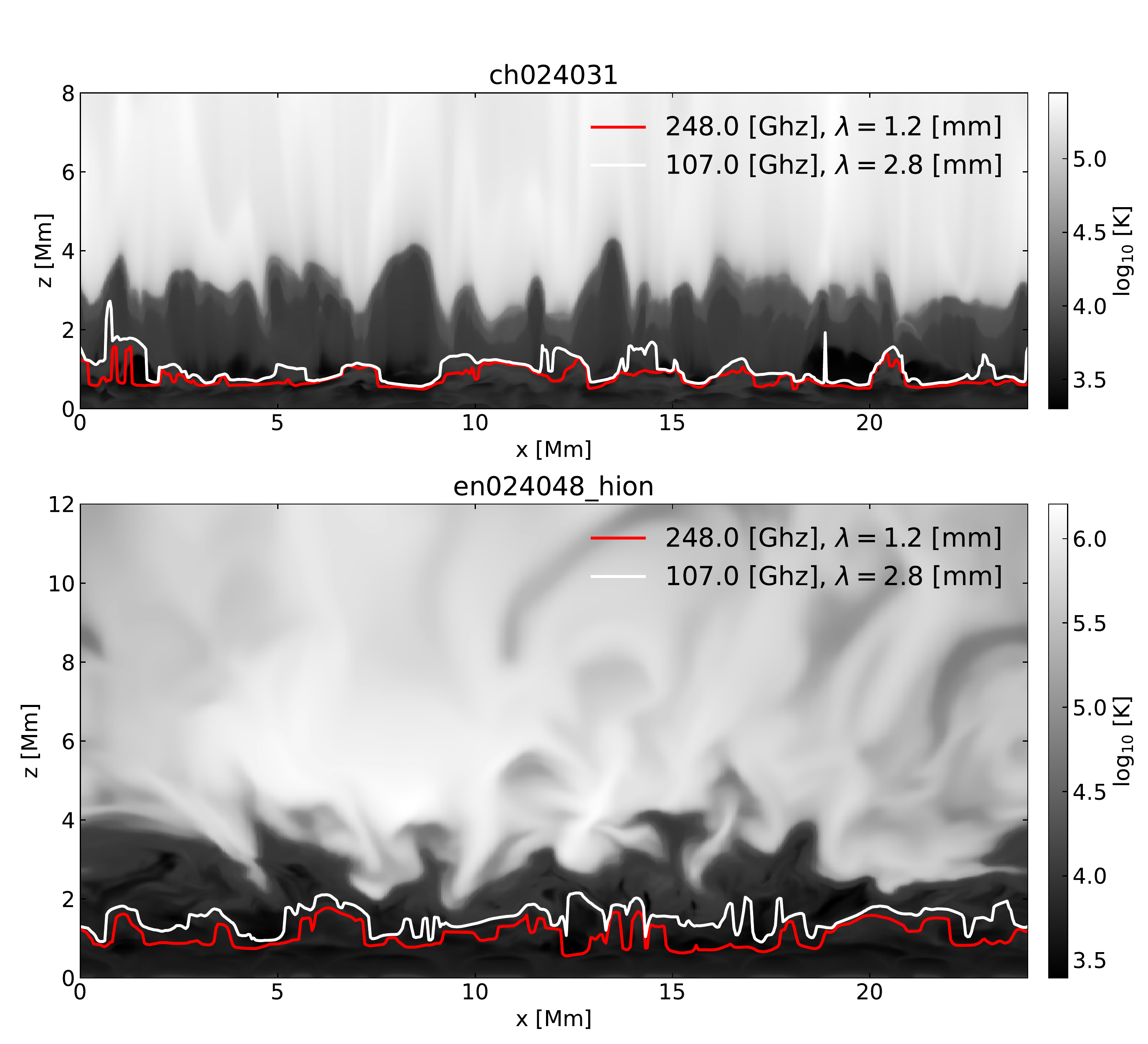} 
    \caption{
    Simulated logarithmic gas temperature (grey-scale) in vertical slices of two 3D rMHD Bifrost simulations: Top: Coronal hole. Bottom: Enhanced network region. The solid lines represent the heights at which the optical depth is unity at a wavelength of 1.2\,mm (red) and 2.8\,mm (white), which correspond to ALMA bands 6 and 3, respectively (see also  Fig.~\ref{fig:bifrost-tau1-over-tg2}). Please note that these heights are only rough proxies of typical formation heights, whereas the contribution functions can have multiple peaks along the line of sight as a result of the complicated thermal structure as can also be seen from the temperature response functions in Fig.~\ref{Fig:RF}.}
	\label{fig:bifrost-tau1-over-tg}
\end{figure*}

\begin{figure*}[t]
	\centering
	\includegraphics[width=0.91\textwidth]{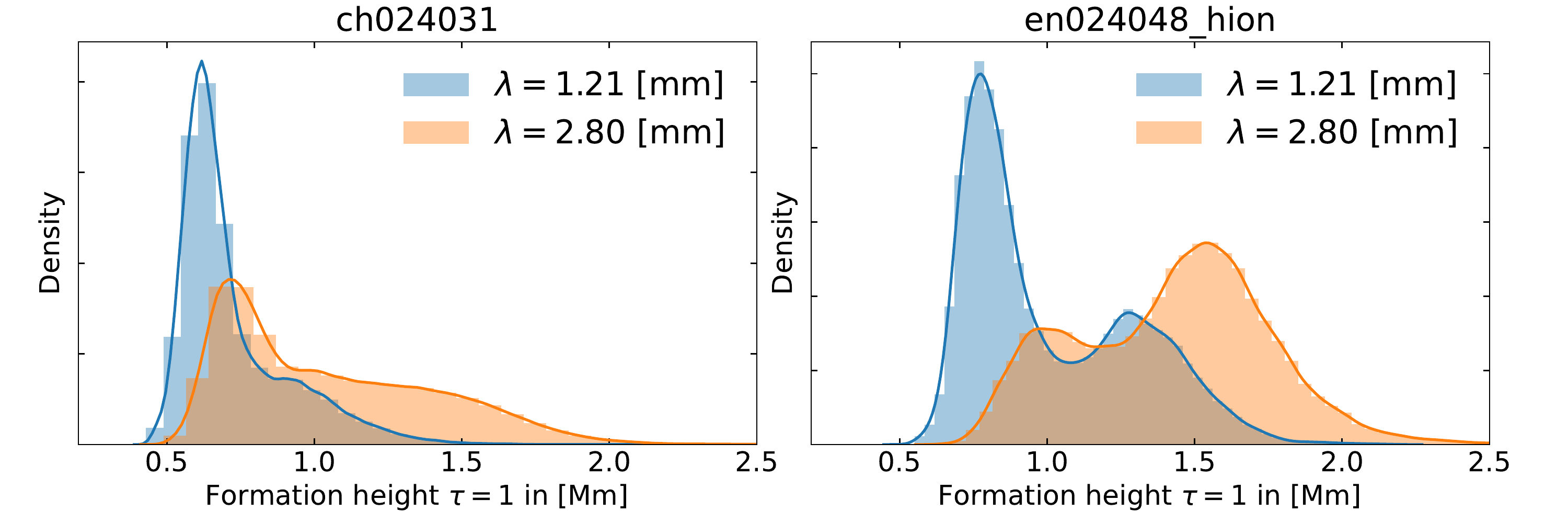} 
    \caption{Distributions of the heights of optical depth unity ($z\,(\tau =1)$) for the two Bifrost models shown in Fig.\,\ref{fig:bifrost-tau1-over-tg} for the continuum at 1.2\,mm (blue) and at  2.8\,mm (orange): Left: Coronal hole. Right: Enhanced network region. }
	\label{fig:bifrost-tau1-over-tg2}
\end{figure*}

Figures\,\ref{fig:bifrost-tau1-over-tg} and \ref{fig:bifrost-tau1-over-tg2} illustrate the formation of the thermal continuum along vertical slices of two snapshots from publicly available\footnote{\url{http://sdc.uio.no/search/simulations}} Bifrost simulations: {\it en\_024048\_hion} -- an enhanced network simulation, including non-equilibrium hydrogen ionisation, and {\it ch024031} -- a coronal hole simulation with LTE ionisation. The heights at which the optical depth ($\tau$) equals one are marked as a proxy for the formation height for simplicity. We note, however, that the contribution functions of the (sub-)mm continuum can potentially be wider and consist of several peaks due to the often complicated structure of the chromosphere leading to multiple optically thin contributions along the line of sight.  
Some ALMA observations indeed show  occasional contributions from several layers \citep{2020A&A...635A..71W}. 
For a detailed insight into the layers that actually contribute to the emission at any given wavelength, these more detailed contribution function should be considered  \citep[e.g.,][]{2015A&A...575A..15L}. The corrugated surfaces of optical depth unity  -- as corrugated as the chromosphere itself -- underscore that the frequencies observed by ALMA probe different heights of the atmospheres at different spatial locations. Therefore, classical semi-empirical models of averaged spectra should be used with caution when interpreting spatially/temporally resolved data.
The histograms in Fig.\,\ref{fig:bifrost-tau1-over-tg2} show that the formation heights ($z\,(\tau =1)$)  are, on average, lower in the coronal hole atmosphere than in the enhanced network. Both the magnetic topology and detailed ionisation balance influence the formation of the millimetre continuum through changes in the electron/proton densities and temperature, directly affecting the free-free opacities. Simulations of more active conditions naturally lead to higher formation heights than in quiet conditions \citep{2020ApJ...891L...8M}. This is true even if non-equilibrium effects are not taken into account because the higher densities in ARs lead to higher (sub-)mm opacities \citep[e.g.,][]{2022A&A...661A..59D}.

Interestingly, even though the Bifrost enhanced network simulation includes much of the relevant chromospheric physics, the averaged synthetic brightness temperatures in {\it en\_024048\_hion} are significantly lower than what ALMA observations suggest. 
\citet{2015A&A...575A..15L} reports $<$$T_{\rm b} (1.25\rm \,mm)$$>$\,$\sim$\,4800\,K and $<T_{\rm b} (3\rm \,mm)>$\,$\sim$\,6100\,K in that simulation, whereas \citet{2017SoPh..292...88W} obtained $<$$T_{\rm b} (1.25\rm \,mm)$$>$\,$\sim$\,5900\,K and $<$$T_{\rm b} (3\rm \,mm)$$>$\,$\sim$\,7300\,K from quiet Sun (QS) observations, values which for the time being are set as reference values for observational ALMA brightness temperatures. \citet{2020A&A...640A..57A} suggests an even higher value: $<$$T_{\rm b} (1.25\rm \,mm)$$>$\,$\sim$\,6300\,K for the QS. We note that it has also been shown that the simulation produces, for UV and visible chromospheric lines, widths and intensities weaker than the observed values in QS regions, which suggests that the simulated chromosphere is overall cooler or/and less dense than the real Sun \citep[e.g.,][]{2013ApJ...772...90L}. As already stated in Sect.~\ref{sec:rmhd}, the inclusion of more physical processes relevant under chromospheric conditions as implemented by \citet{2020ApJ...891L...8M} is an essential step in the right direction, albeit computationally costly.

While there are several codes that can be used for the synthesis of millimetre continuum radiation, the Advanced Radiative Transfer \citep[ART;][]{2021_art} code was developed with the aim to provide a very fast tool for the synthesis of millimetre continuum maps.  
ART uses non-equilibrium electron densities as provided by 3D simulations with Bifrost \citep{2011A&A...531A.154G} and 1D RADYN simulations \citep{1992ApJ...397L..59C} but otherwise assumes LTE conditions. 
The high computational throughput of ART makes it possible to quickly produce time series, even for all individual spectral channels of ALMA instead of just one representative wavelength per (sub-)band. 
ART has been used by \citet{2020A&A...635A..71W,2021A&A...656A..68E,2021RSPTA.37900185E}. 
An example is presented in Fig.~\ref{fig:bifrost-art-sasim} (upper right). 
As a comparison, RH 1.5D \citep{2001ApJ...557..389U,2015A&A...574A...3P} , MULTI \citep{1986UppOR..33.....C} and ART were used to synthesise ALMA Band 3 radiation from a 1D RADYN simulation. The simulation, which was intended as a first experiment towards nano-flares, included a beam of electrons injected during a time interval of 10\,s in an initial atmosphere which is then left to evolve. The differences in brightness temperatures between the results from the three codes, which for this experiment did not contain any non-thermal contributions, agreed within  $\lesssim$\,15\% while the beam was injected and on average less than 6\% during the rest of the time. 

\citet{2021ApJ...914...52F} extended the theory of the free-free emission to the case when the plasma can be composed of many chemical elements in various temperature-dependent ionisation states. The theory takes into account that the plasma, in any resolution element (voxel), can have a distribution of the temperature described by the differential emission measure (DEM), rather than a single value of the temperature. 
\citet{2021ApJ...914...52F} found that taking into account the chemical elements can enhance the free-free emission intensity by up to 30\% depending on the plasma temperature. In the relatively cool chromosphere, where the plasma is only partly ionised, the code takes into account collisions of the free electrons with neutral hydrogen and helium. 
The effect of the magnetic field on the free-free polarisation is also taken into account according to exact magneto-ionic theory, which is important for modelling the free-free emission from the chromosphere \citep{2017A&A...601A..43L}. Another effect controlled by the magnetic field is the gyroresonant emission due to enhanced opacity of the emission at several gyroharmonics---integer multiples of the gyrofrequency (that is proportional to the ambient magnetic field). 
\citet{2021ApJ...914...52F} extended the theory of the gyroresonant emission to the case of multi-temperature plasma, which required to introduce an additional differential density metrics (DDM) complementary to DEM. The code is fully open and available   \citep[see][]{2021zndo...4625572K}. Although the main contribution to the radio emission at frequencies greater than $f\sim100$\,GHz (ALMA Band~3 and above) comes from the chromosphere, coronal contributions are  not negligible in ALMA bands~1 and 2 and the described code is needed to properly model and isolate the coronal contributions. In particular, the gyroresonant contribution can be strong at 34\,GHz and above \citep{2019ApJ...880L..29A}. In addition, the coronal contribution dominates at microwave frequencies, which bear extremely useful complementary information.

\subsubsection{Non-thermal radiation}
\label{sec:rt_nonthermal}

Non-thermal radiation is emitted by non-thermal electrons accelerated in solar flares or relativistic positrons produced due to nuclear interactions \citep{2013PASJ...65S...7F}. The dominant emission mechanism is gyrosynchrotron emission due to spiralling of the charged particles in the ambient magnetic field, while the free-free emission is often important as well (see Fleishman et al. in this special issue). The theory of gyrosynchrotron emission is well developed, but the expressions for the emissivity and absorption coefficients are cumbersome and computationally expensive. To facilitate the computation of gyrosynchrotron emission, \citet{Fl_Kuzn_2010} developed fast gyrosynchrotron codes that programmatically increase the computation speed by orders of magnitude. The  initial versions of the codes permitted only a limited set of analytical distribution functions  of the electrons over energy and pitch-angle to be employed in the computations. 
To facilitate the numerical modelling with a broader range of numerically defined distribution functions, 
\citet{2021ApJ...922..103K} extended the codes for the case of distribution functions defined by numerical arrays. This new version of the code \citep{2021zndo...5139156K} also employs the improved treatment of the free-free emission described in 
Sect.~\ref{sec:rt_thermal}.

\subsection{Simulation of atmospheric seeing and instrumental effects}
\label{sec:degrad}

In order to ensure a fair and meaningful comparison between observations and synthetic observables based on numerical models, 
in principle all effects due to the telescope and instrument, data processing, and Earth's atmosphere must be considered in detail. Typically, for simplicity but often also in lieu of insufficient knowledge of the above mentioned factors, just the (modelled) telescope point spread function (PSF) is applied to synthetic intensity images, resulting in the degradation of the image to the same angular resolution as for the observations. Moreover, such PSFs are often very simplified as contributions such as straylight are usually not known \citep{2009A&A...503..225W}. For such comparisons it should therefore be kept in mind that the degradation effects applies to synthetic images tend to underestimate their effect. 

Properly taking into account instrumental and seeing effects is even more challenging for interferometric imaging as done with ALMA. For solar observations, ALMA's 12-m Array with up to 50 12-m antennas and the Atacama Compact Array (ACA) with up to 12 7-m antennas are correlated, including heterogeneous baselines, resulting in time series or mosaics. At the same time, the up to 4 Total Power (TP) antennas scan the whole disk of the Sun. The obtained visibility data is later calibrated before the brightness temperature maps are reconstructed via an involved deconvolution and imaging process, which is typically based on the CLEAN algorithm 
\citep{1974A&AS...15..417H,2011A&A...532A..71R}. 
The resulting final synthesised beam, which is a byproduct of the CLEAN procedure, has typically an elliptical shape that changes with time as a result of the changing position of the Sun in the sky and thus the corresponding  varying viewing angles of the array. 
While the synthesised beam is equivalent to a PSF and can thus be applied to synthetic intensity maps to match the resolution (see Fig.~\ref{fig:bifrost-art-sasim}, lower right), it is still a simplification of the complicated instrumental setup described above. An even more severe instrumental effect is the sparse sampling of the spatial Fourier space (``$(u, v)$-plane''), which is a direct result of the limited number of  baselines between antennas. This effect is often compensated by following the source as it moves over the sky with time, thus filling different components in the $uv$-space. For the Sun, which changes on short time scales, this so-called rotation synthesis technique is not an option as the small-structure would effectively be blurred out. Instead, brightness temperature maps of the Sun have to be reconstructed solely on the interferometric information retrieved over a very short time (``snapshot imaging''). The cadence of solar observations with ALMA has decreased from 2\,s to only 1\,, which is a great capability for observing the highly dynamic chromosphere but is challenging for imaging based on sparse $uv$-sampling.  

The Solar ALMA Simulator (SASim), which is developed in connection with the ESO-funded ALMA development study \textit{High-cadence Imaging of the Sun} \citep{2019adw..confE..47W}, addresses this problem by aiming at a more realistic simulation of the degradation effects, including the atmosphere above ALMA, interferometric observations and the imaging process. 
As input, SASim uses currently synthetic millimetre brightness temperature maps that are produced with ART from a Bifrost simulation with snapshots output at 1\,s cadence.  
Based on the simobserve task as part of CASA\footnote{Common Astronomy Software Applications (CASA): \texttt{http://casa.nrao.edu}},  the maps are then converted into measurements set similar to what is produced from real ALMA observations of the Sun. In the process, the antenna configuration and the degradation properties of Earth's atmosphere are prescribed. The resulting synthetic measurements set, which is an artificial ALMA  observation of the Sun, is then used as input for the same deconvolution and imaging tool as real observations. For this purpose, the Solar ALMA Pipeline (SoAP) is used \citep{2022A&A...659A..31H,2020A&A...635A..71W}.
See Fig.~\ref{fig:bifrost-art-sasim} (lower left) for an example. 
Such artificial simulation-based observations enable the optimisation of the imaging process as the resulting reconstructed images can be compared to the original (i.e., the SASim input) images for different imaging parameter combinations until the best match between input and output is found. While this procedure is still computationally expensive, it will support the development of new and/or improved observing modes in the future and provide more realistic means for comparing observations with adequately degraded synthetic brightness temperature images.

\section{Advanced reconstruction of model atmospheres from ALMA observations}
\label{sec:processing_analysis}

\subsection{Data inversion}
\label{sec:inv}

\begin{figure*}
	\centering
	\includegraphics[width=0.91\textwidth, trim=0 0.1cm 0 0, clip]{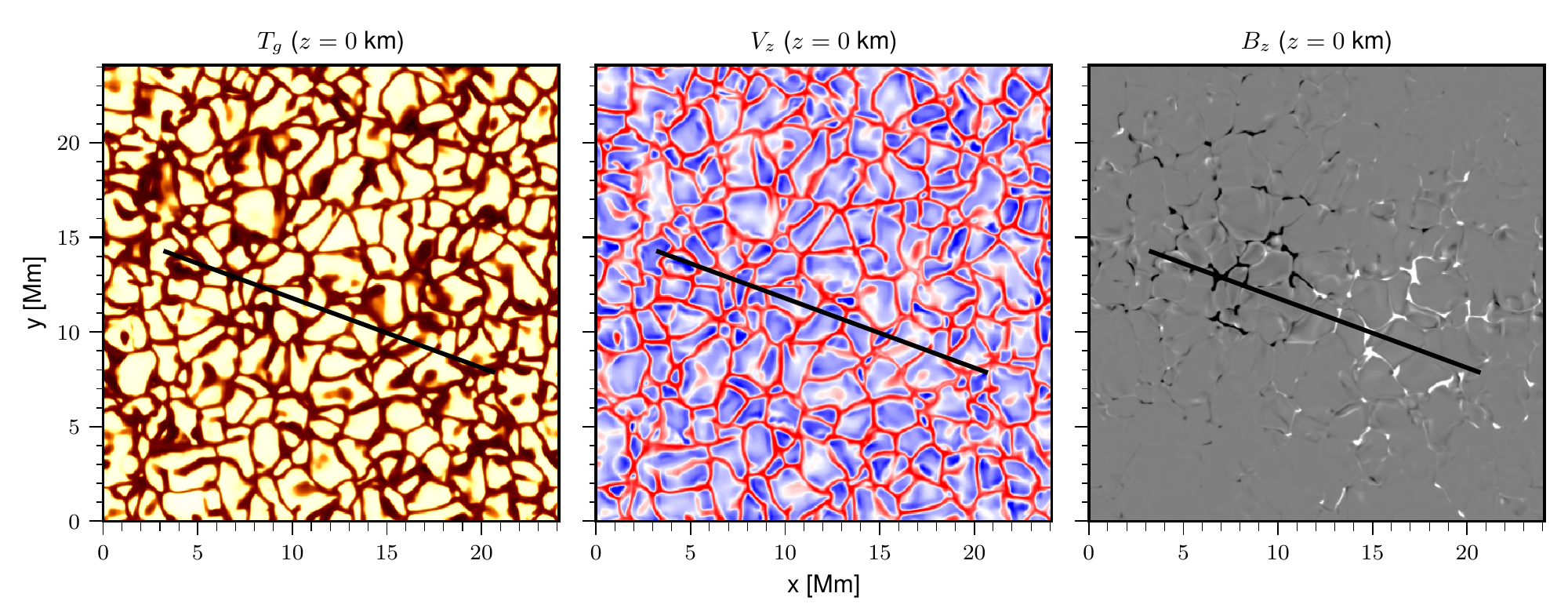}
	\includegraphics[width=\textwidth, trim=0 0.3cm 0 0, clip]{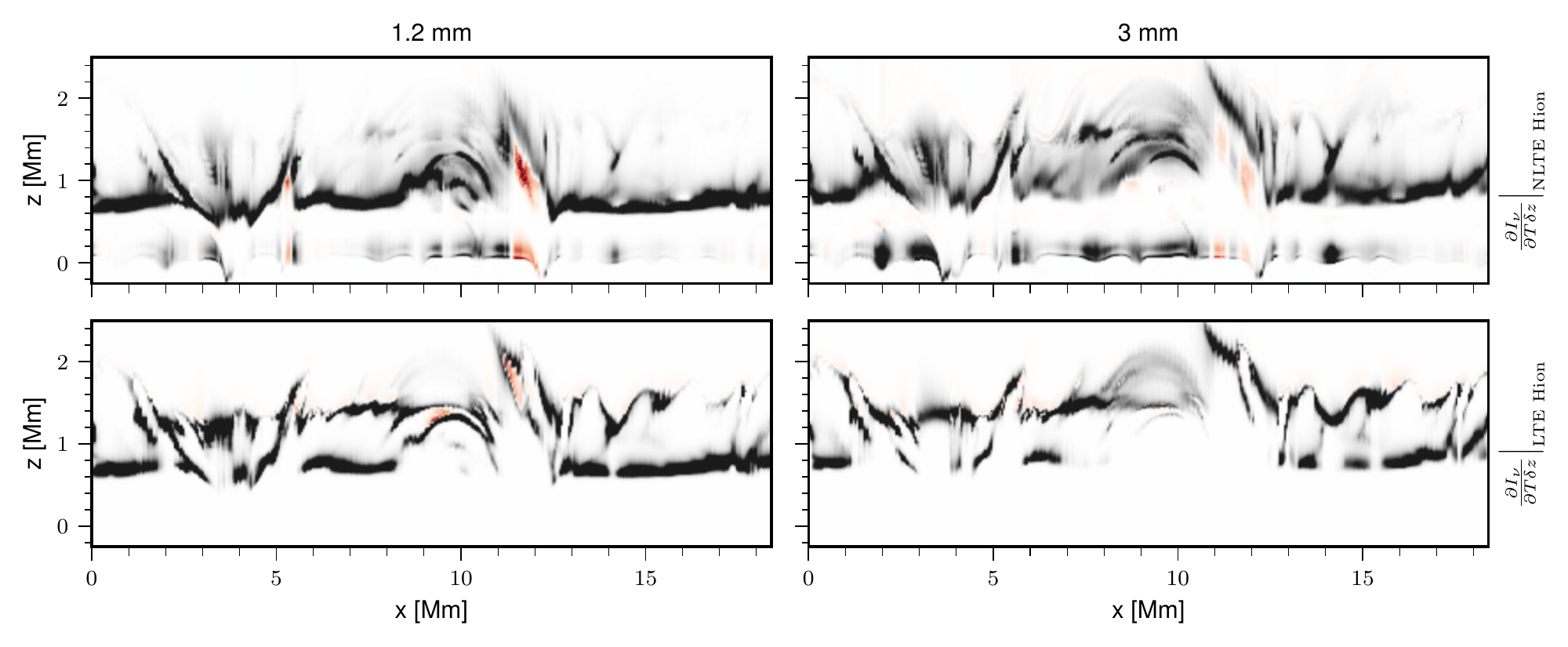}
	\caption{Temperature response function computed with the STiC code at 1.2\,mm and 3\,mm from a vertical slice of a 3D MHD model \citep{2016A&A...585A...4C}. In the middle row the response function has been calculated with a NLTE equation-of-state where the hydrogen population densities are calculated in statistical equilibrium and charge conservation has been imposed in order to calculate the electron density. In the bottom row we have illustrated a LTE case. For context, we have indicated the location of the slice in the upper row over-imposed on the photospheric temperature, the line-of-sight velocity and the vertical component of the magnetic field vector.}
	\label{Fig:RF}
\end{figure*}
Modern inversion techniques allow inferring the physical parameters of model atmospheres from the reconstruction of observed spectropolarimetric data. The model typically includes the stratification of temperature, line-of-sight velocity, microturbulence, and the three components of the magnetic field as a function of column-mass or continuum optical-depth at a reference wavelength. Inversion techniques have been applied systematically to study photospheric datasets with different levels of complexity \citep{1992ApJ...398..375R,2012A&A...548A...5V,2013A&A...549L...4R,2013A&A...553A..63S,2021A&A...656L..20P} under the assumption of LTE. 

The extension of such inversion methods to chromospheric applications was possible by the assumption of statistical equilibrium and plane-parallel geometry in order to calculate the atom population densities \citep{2000ApJ...530..977S}. The most used diagnostic has been the \ion{Ca}{II}~8542~\AA\ line \citep[e.g.,][]{2007ApJ...670..885P,2013A&A...556A.115D,2020A&A...642A.215H}. More recently, the inclusion of lines originating from multiple species and the inclusion of partial redistribution effects has been made possible with the STiC code \citep{2019A&A...623A..74D}, which has enabled the application of inversion methods to datasets including, for example, the \ion{Mg}{II}~h\&k lines \citep[e.g.,][]{2016ApJ...830L..30D,2018ApJ...857...48G,2019ApJ...875L..18S} and the \ion{Ca}{II}~H\&K lines \citep{2019A&A...627A.101V,2019ApJ...870...88E,2022arXiv220301688M}. 

The depth-resolution that can be achieved by inversions greatly depends on the number of spectral diagnostics (lines and continua) included as constraints in the process, as well as their sensitivities to the different regions represented in the model. The inclusion of more diagnostics improves the depth resolution and reduces potential degeneracies in the reconstructed model. However, because of the non-locality of the radiation field in the chromosphere and the consequent decoupling of the source function from the local temperature stratification, NLTE inversions can fail to accurately constrain the temperature stratification in the chromosphere, especially when only one spectral line is included in the inversion.

ALMA millimeter observations provide a powerful diagnostic for chromospheric inversions \citep{2018A&A...620A.124D}. At those wavelengths, the background continuum samples the chromosphere and the source function can be described by the Planck function. However, while the ALMA radiation can be described in LTE, the continuum opacity is set by the electron density and the hydrogen ionisation/recombination balance, which are far from LTE in the chromosphere, and time-dependent NLTE hydrogen ionisation should be included in the calculations \citep[][ see also Sect.~\ref{sec:rt_thermal}]{2006A&A...460..301L,2007A&A...473..625L}. Inversion methods are not sufficiently developed to include time-dependent hydrogen ionisation yet. An alternative approach, which is more accurate than assuming LTE, is to calculate H ionisation in statistical equilibrium and account for this NLTE contribution in the electron densities by imposing charge conservation in each NLTE iteration. The STiC inversion code \citep{2019A&A...623A..74D} allows performing such calculations as part of the spectroscopic inversion process. 
To illustrate the importance of this effect, we have calculated the temperature response function (RF) along a vertical slice from a 3D rMHD simulation (cf.~Sect.~\ref{sec:rmhd}) in LTE, using the proposed statistical equilibrium approach (see Fig.~\ref{Fig:RF}). The response function is representative of the sensitivity of each grid cell to temperature perturbations. 
For both receiver bands, the LTE response function has significant peaks around heights of  0.8 and 1.2\,Mm. While the peak at the lower height tends to be more prominent for Band~6 than compared to Band~3, there are many locations for both bands where peaks at both heights occur, resulting in bimodal RFs. 
When NLTE H ionisation is included in the calculation, the response function generally becomes less localised, basically extending across a height range from as high as 1.8-2.0\,Mm to as low as $\sim 0.1$\,Mm for both receiver bands. 
Within this range, also the NLTE RF often shows multiple peaks. 
The computationally easier LTE approach is clearly less accurate particularly in cold atmospheric pockets after the passage of a shock wave, where a full time-dependent treatment would allow for higher ionisation fractions than those predicted by statistical equilibrium. A detailed treatment of time-dependent hydrogen ionisation is therefore essential for modelling the chromospheric  emission at mm wavelengths. 

\begin{figure}[t]
    \centering
    \includegraphics[width=\linewidth]{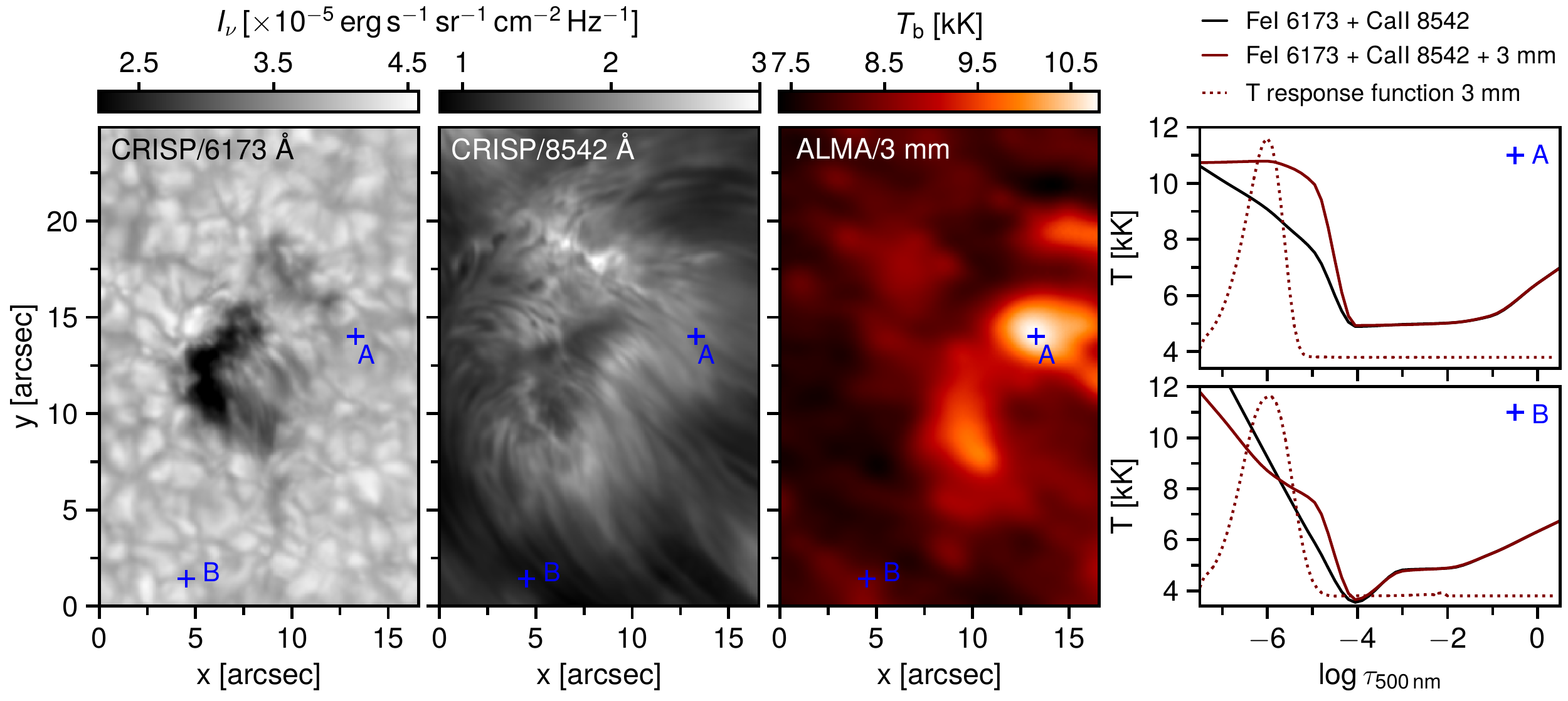}
    \caption{NLTE inversions of SST/CRISP and ALMA observations of an active region \citep{2022A&A...661A..59D}. {\it From the left to the right:} intensities in the continuum at 6173\,\AA~and in the core of the Ca\,II 8542\,\AA~line, ALMA Band~3 brightness temperature map, temperature as function of logarithmic optical depth (solid lines) derived from CRISP and CRISP+ALMA inversions at the locations {\it A} and {\it B}, and temperature response functions at 3\,mm in arbitrary units (dotted lines) for the CRISP+ALMA models.}
    \label{fig:inv}
\end{figure}

While it is not possible to constrain all the inversion parameters using ALMA continuum observations alone, the combination of ALMA data with spectroscopic/spectropolarimetric datasets from other facilities such as DKIST, the 1-m Swedish Solar Telescope (SST), BBSO, or NASA's IRIS satellite is extremely powerful in inferring atmospheric models. For example, \citet{2020A&A...634A..56D} utilised datasets from IRIS and ALMA to constrain the average value of microturbulence in plage down to approximately 5~km\,s$^{-1}$ within the range of optical depths probed by the \ion{Mg}{II} h and k lines ($\log\tau_{\rm 500\,nm}=[-6,-4]$) using STiC. Furthermore, \citet{2022A&A...661A..59D} used co-temporal SST/CRISP and ALMA observations to reconstruct a model atmosphere of an AR undergoing magnetic reconnection (Fig. \ref{fig:inv}). The rightmost panels in Fig. \ref{fig:inv} illustrate that the addition of the continuum intensities at 3\,mm to NLTE inversions of the \ion{Fe}{i} 6173\,\AA~and \ion{Ca}{ii} 8542\,\AA~lines observed by CRISP has a significant impact on the inverted temperatures in the chromosphere ($\log \tau_{\rm 500\,nm}$\,$<$\,$-4$) where the temperature response function at 3\,mm peaks, and thus provides more accurate models for, in this case, comparisons with rMHD simulations of magnetic flux emergence into the chromosphere. 
\citet{2022arXiv220508760H} investigated inversion strategies for co-temporal ALMA and IBIS observations in the \ion{Na}{i} 5896\,\AA~and \ion{Ca}{ii} 8542\,\AA~lines using STiC and showed that performing the inversions on a column-mass scale, instead of an optical depth scale, helps to resolve the chromosphere, while the inclusion of the NLTE electron densities as described above leads to a warmer chromosphere (by $\sim$1,000\,K) compared to the LTE case. ALMA Band~3 continuum measurements were successfully inverted together with the spectral lines, but the Band~6 temperatures were difficult to reproduce. The latter could be due to the multi-modal contribution function at 1.2\,mm \citep{2015A&A...575A..15L} reproduced in Fig.\,\ref{fig:bifrost-tau1-over-tg2}, introducing uncertainties in the determination of the height of the observed emission, and the time-dependent opacity effects (mostly as a result of the variations of the electron density, see Sect.\,\ref{sec:rt_thermal}) not accounted for in the code. 

Inverse modelling of ALMA brightness temperature maps along with spectroscopic or spectropolarimetric observations at higher frequencies entails several challenges in the data acquisition and analysis, such as the non-simultaneity of the observations at different wavelengths, significant differences in spatial resolution, the uncertainties in the absolute flux calibration, and the simplifying assumptions in inversion codes, namely hydrostatic equilibrium and 1D radiative transfer. Therefore, improving data inversions is on-going work with large scientific potential but also requires further coordinated campaigns between ALMA and other observatories in order to produce the needed data sets.

\subsection{Magnetic field reconstructions}

Probing and modelling coronal magnetic field is a key challenge in the solar physics. Measurements of the coronal magnetic field are difficult and yet very rare \citep[e.g.,][]{2004ApJ...613L.177L,2006ApJ...641L..69B,2014A&ARv..22...78W,2016ApJ...833....5S,2017SSRv..210..145C,2019ApJ...874..126K,2020Sci...367..278F}, giving  modelling of the coronal magnetic field  an even more important role. The most easily accessible, routine approach to this modelling is the use of nonlinear force-free field (NLFFF) reconstruction \citep[e.g.,][]{2012LRSP....9....5W,2014A&ARv..22...78W}.  Typically, such reconstructions are initiated with a vector boundary condition derived from a routine Zeeman diagnostics at photospheric heights. 
A challenge associated with this boundary condition is that the magnetic field is not necessarily in a force-free state in the photosphere, which is in conflict with the assumption that the field is force-free. 

\citet{2019ApJ...870..101F} developed an advanced NLFFF code that can use additional (to the photospheric ones) magnetic constraints at various locations and various heights in the chromosphere and/or corona while performing the NLFFF reconstruction. The quality of the NLFFF reconstruction increases notable the more  complete the set of additional constraints is. 
In this regard, multifrequency imaging spectropolarimetry in various ALMA spectral bands would enable measurements of the LOS magnetic field component at different chromospheric heights \citep{2017A&A...601A..43L}. Adding these measurements to the NLFFF reconstruction code promises a much improved magnetic model of a given AR. 
The necessary observing modes and post-processing steps are still under development but might be offered in the coming years, possibly starting with full polarisation observations of the Sun in Band~3 from as early as Cycle~10 (i.e. 2023-4). 

\section{Examples of scientific applications}
\label{sec:examples} 
\subsection{Atmospheric stratification and centre-to-limb variation}
\label{sec:clv}

When observing the solar atmosphere not directly from the top but at an inclined viewing angle (i.e. $\mu = \cos \theta < 1$), the line of sight traverses a longer path through the solar atmosphere for a given height difference, resulting in optical depth unity being reached higher up in the atmosphere. 
As the change in viewing angle corresponds to the variation from observations at different distances from the disk-centre ($\mu = 1$) towards the limb  ($\mu = 0$), observations of the centre-to-limb variation of the continuum emission contain information about the atmospheric stratification (i.e. the height dependence of the underlying plasma properties) and can thus  be used to constrain models of the solar atmosphere \citep[see, e.g.,][]{2018CEAB...42....1B}. \citet{2017A&A...605A..78A,2020A&A...640A..57A} used ALMA brightness temperatures maps in the range $\sim$1-3\,mm at different positions on the solar disk to infer a relation between kinetic temperature and optical depth, which was shown to be consistent with the widely used FALC QS model \citep{1993ApJ...406..319F} within the uncertainties. Using ALMA observations, \citet{2019SoPh..294..163S} determined lower limits for the limb brightening effect on the order of 10\% in Band~3 and 15\% in Band~6. These results agree with the range of values independently derived by \citet{2022MNRAS.511..877M}. 
Such empirically derived model atmospheres provide crucial tests for numerical models (Sect.~\ref{sec:rmhd}) for which the centre-to-limb variation of the (sub-)mm can be computed by repeating the radiative transfer calculations (Sect.~\ref{sec:rt}) under varying viewing angles.

\subsection{Small-scale dynamics in Quiet Sun regions}

As established by observations using other chromospheric diagnostics such as the \ion{Ca}{II}\,K line, internetwork QS regions exhibit a highly dynamic mesh-like pattern produced by the interaction of ubiquitous propagating shock waves \citep[e.g.,][]{2006A&A...459L...9W}, unless obscured by overlying fibrilar structures outlining the magnetic field. 
The expected appearance of these signatures of chromospheric shock waves at (sub-)mm wavelengths had been simulated based on 1D simulations \citep{2004A&A...419..747L} and 3D simulations \citep{2005ESASP.560.1035W,2007A&A...471..977W} already before the advent of ALMA. 
The simulations confirmed that sufficient spatial resolution, which is particularly challenging at (sub-)mm wavelengths, and a high temporal cadence are essential for enabling the detection of shock waves. However, also the viewing angle and thus the location on the solar disk impact the detectability of shock waves due to the potential   superposition of structures at different heights along inclined lines of sight (cf. Sect.~\ref{sec:clv}). 
As a consequence, ALMA is expected to observe the ubiquitous chromospheric shocks in the QS with higher contrast at disk centre than closer to the limb \citep{2007A&A...471..977W}. 

As illustrated in Fig.~\ref{fig:bifrost-tau1-over-tg} for a Bifrost simulation, the formation height of the millimetre continuum varies significantly when the atmosphere is buffeted by shock waves. Consequently, the brightness temperature  increases  temporarily by up to several thousand kelvin, as long as the shock structure remains optically thick at a given wavelength \citep{2021RSPTA.37900185E}. 
Comparable ALMA observations show bright small scale features in the QS with typical lifetimes of $\sim$\,1-2\,min and amplitudes of a few hundred kelvin above background \citep{2020A&A...644A.152E, 2021A&A...652A..92N}. 
Taking into account the impact of ALMA's limited spatial resolution on the appearance of small scale structures and the resulting reduction of the amplitudes of the dynamic bright features, the detected bright features are consistent with upwardly propagating shocks as predicted by numerical simulations \citep{2020A&A...644A.152E}.
See Fig.~\ref{fig:bifrost-art-sasim} for an illustration of the shock-induced chromospheric pattern before and after applying the effects of limited spatial resolution to simulated millimetre continuum images. 
The amplitude reduction of bright features in the Bifrost simulation at wavelengths corresponding to all ALMA receiver bands\,3\,-\,10 (0.34\,mm - 3.22\,mm) and corresponding spatial resolutions with different ALMA antenna configurations was studied by \cite{2021A&A...656A..68E}, providing  conversion factors that could in principle be used to correct for the amplitudes of bright features in observational data on a statistical basis although with large uncertainties. 
Furthermore, propagating shocks give rise to steep temperature gradients in the atmosphere. As numerical simulations show, the slope of the brightness temperature continuum can in principle be used to identify signatures of shocks and also the potential existence of multiple shock components  propagating at different velocities as shown by  \cite{2021RSPTA.37900185E}. The diagnostic potential of splitting ALMA data sets into sub-bands has recently been shown for the study of propagating transverse waves by \citet{2022A&A...665L...2G} and for propagating shock waves by \citet{2022A&A..sub.ALMA.SB.E}. 

Shocks may play a significant role in the heating of the quiet chromosphere, but addressing this science case with ALMA beyond pure detection requires follow-up observations with higher spatial resolution than the ones available so far. 
The use of wider array configurations in combination with improved imaging routines has the potential to reach a resolution of a few 0.1'' for ALMA's higher receiver bands. Determining the exact requirements could be achieved through  the modelling approach illustrated in Fig.~\ref{fig:bifrost-art-sasim}. 
In addition, studying the propagation of shock waves through the (sub-)mm continuum forming layers by means of the methods introduced in Sects.~\ref{sec:forwardmodelling} and \ref{sec:processing_analysis), which includes in particular} inversions and detailed (statistical) comparisons of high-quality ALMA observations and artificial observations based on state-of-the art numerical 3D simulations,  promises essential insights into the formation heights of the (sub-)mm continuum as function of wavelength and the atmospheric stratification for different regions on the Sun. 
Please see  \citet{2020A&A...634A..56D} and \citet{2020A&A...644A.152E,2021A&A...656A..68E} for recent studies employing these methods.

\subsection{Active Regions and Flares} 
\label{sec:arnflare}

Well-developed ALMA diagnostics could contribute significantly to progressing our understanding of the thermal and magnetic properties of ARs and flares. 
Modelling the magnetic topology of a steady-state AR is now routinely possible based on NLFFF reconstructions constrained by routinely available photospheric vector magnetograms \citep{Fl_etal_2015ALMA,2021ApJ...909...89F}. At the same time, dynamic phenomena are particularly hard to reproduce in numerical simulations for several reasons: 
The time scales that need to be resolved span the evolution of ARs (from hours to days) to the rapid changes that occur during flares (i.e., to the fractions of seconds). The spatial scales likewise span a large range of values from the large extent of ARs (up to several 10\,Mm) to the small computational cell size that is needed to adequately model all relevant processes (on the order of km towards sub-grid modelling at even much smaller scales). Finally, the presence of stronger than average magnetic field,  the occurrence of magnetic reconnection across different scales,  and the resulting impact on the chromospheric dynamics and, here of particular relevance, the emission at (sub-)mm wavelengths pose particularly challenging demands on numerical modelling (cf. Sect.~\ref{sec:forwardmodelling}). Consequently, forward modelling of the (sub-)mm continuum intensities emerging from semi-empirical models is therefore a natural starting point and, despite the obvious limitations, a complementary way of gaining insight into the physics of AR phenomena through detailed comparisons between synthetic and observed brightness temperatures.  
At the moment of writing, there are a few examples of the application of this methodology in the literature.
For instance, \citet{2017ApJ...850...35L} compared brightness temperature measurements of a sunspot at two different wavelengths with the expected values computed from several 1D static, semi-empirical models in the literature and concluded that none of the models provided a good fit for the penumbra. This means that ALMA can provide constraints on the temperature gradient in those structures. Semi-empirical models have also been used to investigate the observed correlation between the width of the \ion{H}{$\alpha$} line and the brightness temperatures at 3\,mm in plage -- which was explained based on the mutual dependence of both diagnostics on the hydrogen atom level populations
\citep{2019ApJ...881...99M}. Improving the agreement between empirical models and ALMA observations can also be done in an automatic or iterative way using NLTE inversion codes (Sect.~\ref{sec:inv}) when co-temporal spectroscopic observations at optical or ultraviolet (UV) wavelengths are likewise available \citep{2020A&A...634A..56D,2022A&A...661A..59D,2022arXiv220508760H,2022FrASS...9.1118D}.  
Moreover, one-dimensional time-dependent simulations have been used to study the response of the solar atmosphere to flares. For instance,  \citet{2017A&A...605A.125S} used RADYN simulations to evaluate the diagnostic potential of flare emission in the thermal infrared and sub-millimetre range.

As an example of the applications of multi-dimensional numerical models in the context of interpreting ALMA observations of the Sun, the comparison between ALMA Band~6 observations of AR plage supported by IRIS UV spectroscopy and synthetic spectra from a 2.5D Bifrost simulation suggests that an observed bright linear feature emanating from the plage region is an on-disk type-II spicule, confirming their multithermal nature \citep{2021ApJ...906...82C}, while other small-scale plage brightenings at 1.25\,mm could be signatures of magnetoacoustic shocks \citep{2021ApJ...906...83C}. ALMA Band~3 observations of the same plage region also show out-of-phase oscillations in brightness temperature, feature size, and horizontal velocities, which are evidence of sausage-mode and kink-mode waves \citep{2021RSPTA.37900184G}.
AR transients akin to UV-bursts \citep{2018SSRv..214..120Y} with nanoflare-like energies have been observed in a flux emergence region with Band~3, and their characteristics agree with synthetic thermal emission from a 3D Bifrost simulation of flux emergence, showing several magnetic reconnection events \citep{2020A&A...643A..41D}. 
For a microflare detected at 3\,mm in coordination with IRIS and Hinode/XRT by \citet{2021ApJ...922..113S}, the timing of the millimetre brightening with respect to the soft x-rays indicated that the upper chromosphere was being heated by non-thermal particles in this event. Such an effect has also been pointed out by 
\citet{2019SoPh..294..150V} 
using lower resolution observations at 1.4\,mm with the Solar Submillimeter Telescope. 
Please note that the formation of the millimetre continuum emission under flaring conditions, similarly to what has been done for quiet conditions (Sect.~\ref{sec:rt_thermal}), 
and the interpretation of those diagnostics have been addressed previously within the given instrumental capabilities \citep[see, e.g.,][]{1992SoPh..140..139S,1993AdSpR..13i.289K,1996ApJ...458L..49S,2013A&ARv..21...58K}. 

Finally, full 3D modelling of ARs and flares requires a 3D model of the magnetic structure to start from, which can be obtained from a NLFFF reconstruction or other means. To make the modelling with such 3D data cubes realistic, the magnetic ``skeleton'' must be supplied by the thermal plasma, non-thermal particles (in case of flares), placed at the right location at the Sun, and supplied by various means for emission computation and meaningful  data-to-model comparison. 
\citet{Nita_etal_2015,Nita_etal_2018} developed a powerful dedicated tool, called GX Simulator, that is designed specifically for the goals outlined above. More than that, the GX Simulator includes an integrated mean to create a 3D model, an automated model production pipeline (AMPP), with minimal input from the user: Only the date and time of the event have to be defined along with the anticipated resolution and sizes of the data cube. Once specified, AMPP will download all required magnetic maps obtained with SDO/HMI and other context data, perform magnetic reconstruction as well as several other steps that are needed to create a 3D model. Then, magnetic flux tube(s) can be interactively created and designated to serve as the model flare loop(s). The user can control properties of the thermal and non-thermal components in these flaring loops, compute mm/radio, EUV, and X-ray emissions from the model, and compare the synthetic images and spectra with the observed ones, in particular, with ALMA data.

As outlined above, there are many ways how ALMA observations of flares and corresponding numerical models could complement each other in order to advance our understanding of this intricate process that is essential for the heating of the solar atmosphere over a large range of scales from nano-flares to X-class flares and, for other stars than the Sun, even towards mega-flares. 
Owing to its complexity and still insufficient knowledge regarding central aspects, many simulations necessarily focus on particular parts of the flare phenomenon and/or have to make simplifying assumptions. There are, however, efforts to capture flares self-consistently in full 3D time-dependent simulations as those as attempted with the MURaM simulations by \citet{2019NatAs...3..160C}. Such simulations could be used for calculating the corresponding (sub-)mm emission as it could be observed with ALMA, though they do not include non-thermal particles responsible for a good portion of the flare emission in the ALMA bands yet.  The comparison of both would then, as for the other examples described in this article, allow to identify potential shortcomings of the models, which is necessary for future improvements towards a more detailed and complete picture. 
ALMA observations of flare emission, either with the already available continuum mapping capabilities and even more so with future full polarisation capabilities, would provide valuable information complementary to existing flare diagnostics. 
Unfortunately, to the authors' knowledge at the time of writing, there are no good examples of flare observations with ALMA yet, maybe except for one microflare reported by \citet{2021ApJ...922..113S}. 
Please refer to Fleishman et al. in this issue for more information about how ALMA observations could advance our understanding of solar flares.

\subsection{Solar prominences}
\label{sec:prominence}

In the field of prominence research, numerical simulations of (sub-)mm radiation observed by ALMA can help in two areas. The first is the visibility of large and small-scale structures of prominences and filaments in the ALMA observations and the connection between the structures observed at (sub-)mm wavelengths and in other parts of the spectrum. Methods that generate such simulated ALMA observations can be based on actual data observed at different wavelengths -- see, e.g. the simulations of the prominence fine structure visibility based on H$\alpha$ observations performed by \citet{2015SoPh..290.1981H} -- 
or on the synthesis of the radiation at (sub-)mm wavelengths emerging from models of prominence plasma (cf.~Sect.~\ref{sec:forwardmodelling}). 
The first synthetic ALMA-like imaging of simulated prominences was done by \citet{2016ApJ...833..141G}. These simulated ALMA data represent the way in which the 3D Whole-Prominence Fine Structure (WFPS) model of \citet{2015ApJ...803...64G} would appear in different ALMA bands. 
This model combines realistic 3D magnetic field configurations with a detailed description of the prominence plasma located in the magnetic dips. In doing so, the 3D WPFS model emulates entire prominences with their numerous fine structures and offers high spatial resolution reaching the 
best potential capabilities of ALMA. However, we should note that the current ALMA observations of prominences \citep{2022ApJ...927L..29H,2022MNRAS.513L..30L} or filaments \citep[][in this special issue]{fspas.2022.898115} do not reach such a resolution. This is because ALMA has not achieved yet its full potential for solar observations and offers only more compact antenna configurations for solar targets. For more details on the actual and simulated observations of prominences and the predictions of the visibility of prominence fine structures, see the review by \citet[][]{10.3389/fspas.2022.983707} in this special issue. 

The second area where numerical simulations can help the prominence research is the understanding of the potential of ALMA observations as diagnostics for the kinetic temperature of the solar plasma. It has been known for some time that a measure of the kinetic temperature of the observed plasma can be derived from simultaneous observations in at least two different (sub-)mm wavelengths \citep[see, e.g.,][]{2004A&A...419..747L,2012SoPh..277...31H,2015SoPh..290.1981H}. However, only the recent modelling efforts demonstrated that such derivation is not entirely straightforward and that in order to achieve sufficient precision,  the observed plasma needs to be optically thick at one of those wavelengths and optically thin at the other. 
It must be noted here that such observations would require a new capability that ALMA is currently not offering and likely will not offer in the near future, namely simultaneous solar observing with two sub-arrays, each in a different receiver band.    
In the absence of such simultaneous multi-band ALMA observations, thermal properties of the observed plasma can be derived from a combination of ALMA observations and observations in other wavelength ranges, for example in the H$\alpha$ line \citep[see][]{2022ApJ...927L..29H,2022MNRAS.513L..30L}.

\begin{figure*}
	\centerline{\includegraphics[width=15cm]{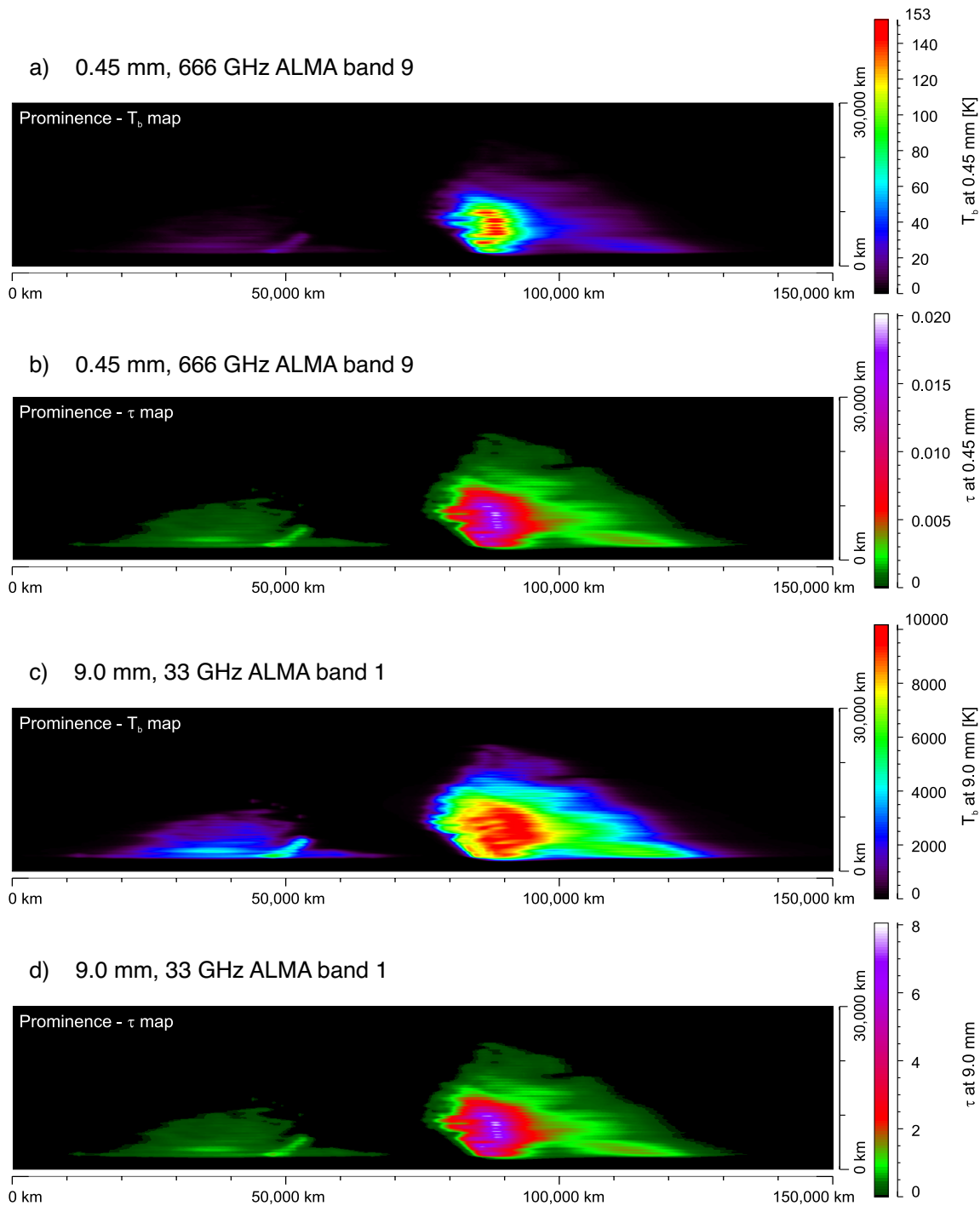}}
	\caption{Synthetic brightness temperature maps (panels a and c) and optical thickness maps (panels b and d) showing the 3D Whole-Prominence Fine Structure model of \citet{2015ApJ...803...64G} in a prominence view. Top two panels were obtained at the wavelength of 0.45\,mm, which corresponds to the 666\,GHz ALMA Band~9. The bottom two panels were obtained at the wavelength of 9.0\,mm (33\,GHz ALMA Band~1). Displayed color scales are unique for each panel. Adapted from \citet{2016ApJ...833..141G}.}
	\label{Fig:Gunar2016}
\end{figure*}

The first study of ALMA's diagnostic potential for determining the properties of prominence plasma was done by \citet{2017SoPh..292..130R}. These authors used 2D cylindrical prominence models by  \citet{2009A&A...503..663G}, both in the iso-thermal iso-baric configuration and in the configuration incorporating the Prominence-Corona Transition Region (PCTR). The study of \citet{2017SoPh..292..130R} shows that the realistic assumption of the presence of multi-thermal plasma along a line of sight leads to uncertainties in the determination of the kinetic temperature of the observed plasma. The authors then demonstrate that such uncertainties can be minimized by using ALMA observations at longer wavelengths (e.g., at 9 mm). At such wavelengths, the observed prominence plasma can be expected to be mostly optically thick, leading to reliable diagnostics of its thermal properties. 

The relationship between the observed brightness temperature in various ALMA bands and the actual kinetic temperature of the observed plasma was further investigated by \citet{2018ApJ...853...21G}. To do so, these authors used the simulated ALMA observations of \citet{2016ApJ...833..141G} at wavelengths at  which the prominence plasma is optically thin (0.45\,mm, corresponding to 666\,GHz in Band~9) and optically thick (9.0\,mm, corresponding to 33\,GHz in  Band~1). The resulting brightness temperature maps and the optical thickness maps are shown in Fig.~\ref{Fig:Gunar2016}. 
The 0.45\,mm and 9.0\,mm wavelengths were chosen to assure that the studied plasma is completely optically thin (at 0.45\,mm) and at the same time that a significant portion of the analysed prominence is optically thick (at 9.0\,mm). However, we should note that at the time of writing solar observations with ALMA are offered only in bands 3, 5, 6, and 7, while bands~9 and 1 are not available yet. 
We also note that the current status of the development of Band~1 indicates that it will reach a slightly shorter wavelength of 8.6\,mm (35\,GHz) instead of the 9.0\,mm (33\,GHz) used here. Moreover, ALMA Band~10 will provide access to even shorter wavelengths than those covered by Band~9 assumed here. 

The results of \citet{2018ApJ...853...21G} confirm that the brightness temperature simultaneously measured by ALMA in the above-specified bands can be used to derive the kinetic temperature in every pixel of the observed prominences, for example using the method described in Sect.~3 of \citet{2018ApJ...853...21G}. However, not all of the derived values of kinetic temperature are accurately representing the thermal conditions of the plasma distributed along lines of sight passing through individual pixels because the observed prominence cannot be generally assumed to be optically thick (i.e. have optical thickness above unity) in all pixels, even at the 9.0\,mm wavelength. This can be clearly seen in the optical thickness map in panel~d of Fig.~\ref{Fig:Gunar2016}, where large parts of the modelled prominence have an optical depth of $\tau_{9.0\mathrm{mm}}<1$. Therefore, without the added information about the actual optical thickness, which is in this case provided by the model, the derived values of the kinetic temperature cannot be automatically assumed to be accurate in all pixels. This is because the basic approximation of the method used to derive the kinetic temperature is not valid if the plasma is not optically thick. Thanks to the use of numerical simulations, \citet{2018ApJ...853...21G} could determine that the kinetic temperature values obtained in pixels with an optical thickness larger than 2 typically have an accuracy better than 1000\,K compared to the true kinetic temperature of the studied plasma. Such a study was possible because the authors used realistic numerical simulations, demonstrating the essential value of simulations for testing adequate methods for the interpretation of observations.

The 3D WPFS model of \citet{2015ApJ...803...64G} contains detailed information about the properties of the prominence plasma including its kinetic temperature. Thanks to the (sub-)mm radiation synthesis method  described in \citet{2016ApJ...833..141G}, the authors could synthesise the specific intensities at any wavelength along any LOS crossing the simulated prominence. That led to an ideal set of co-spatial simulated observations (panels a and c of Fig.~\ref{Fig:Gunar2016}) coupled with the implicit knowledge of the corresponding optical thickness (panels b and d of Fig.~\ref{Fig:Gunar2016}). The major problem lies in the fact that the optical thickness values are not explicitly known based on ALMA observations alone -- but can be derived from other observations, such as those in the H$\alpha$ line \citep[see][]{2015SoPh..290.1981H,2022ApJ...927L..29H,2022MNRAS.513L..30L}. However, \citet{2018ApJ...853...21G} showed that thanks to the use of numerical simulations it is possible to determine a criterium for the minimum value of the measured brightness temperature in the optically thin wavelengths (in this case at 0.45\,mm) above which the optical thickness at the optically thick wavelengths is with great confidence higher than a required value.  
The second part of the study conducted by \cite{2018ApJ...853...21G} showed that the values of the kinetic temperature derived from the ALMA observations correspond to the mean kinetic temperature of the observed plasma weighted by the contribution function of the emission in the optically thick wavelengths. For more details see Sections~5 to 7 of \citet{2018ApJ...853...21G}.

\subsection{Coronal rain}

Besides the large prominence structures discussed in Sect.~\ref{sec:prominence}, condensation phenomena occur far more frequently in the solar corona in the form of coronal rain. This phenomenon, which is also thought to be the seeds leading to prominences, corresponds to cool ($10^3-10^5$~K) and dense ($10^{10}-10^{12}$~cm$^{-3}$) plasma occurring in a timescale of minutes seemingly condensing out of nowhere at coronal heights, and flowing predominantly downwards along loop-like trajectories under the action of gravity and gas pressure mainly \citep{Kawaguchi_1970PASJ...22..405K,Leroy_1972SoPh...25..413L,Foukal_1978ApJ...223.1046F, Antolin_Rouppe_2012ApJ...745..152A}. The leading explanation for coronal rain is thermal instability (TI) within a structure in thermal non-equilibrium and therefore reflects mostly the MHD thermal mode becoming unstable and generating a condensation from the localised loss of pressure \citep{Antiochos_1999ApJ...512..985A, Antolin_2020PPCF...62a4016A}. Numerical simulations have shown that the easiest path towards thermal instability within a structure such as a coronal loop is to have strongly stratified and high-frequency heating \citep{Klimchuk_2019ApJ...884...68K}. This state, known as thermal non-equilibrium (TNE), leads to TNE cycles for heating that remains largely unchanged over time. The details of this process and the various observable signatures can be found in \citet{Antolin_Froment_10.3389/fspas.2022.820116}.

\begin{figure*}
	\centerline{\includegraphics[width=15cm]{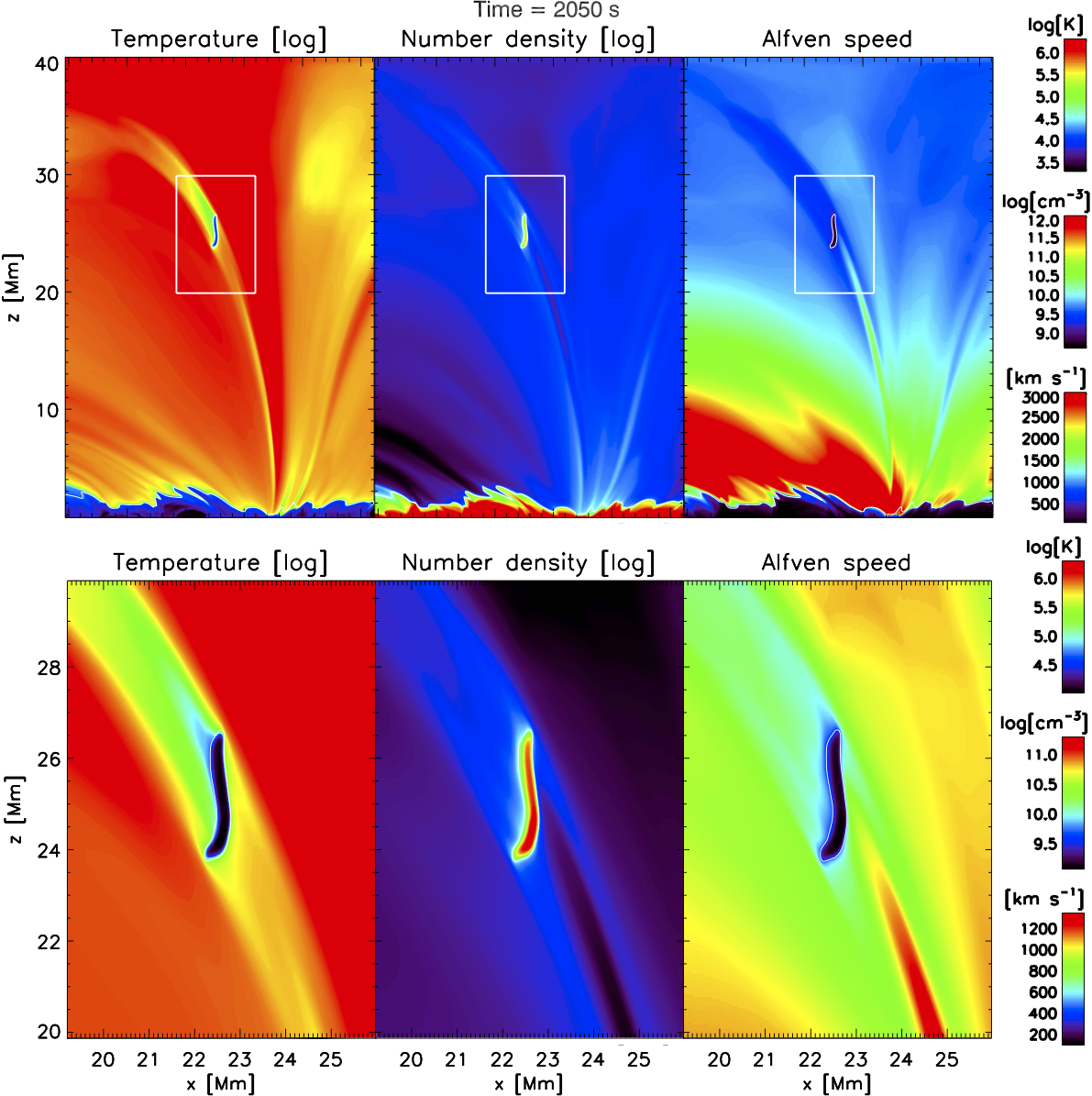}}
	\caption{A coronal rain event self-consistently produced in a 2.5D rMHD simulation with Bifrost (Sect.~\ref{sec:rmhd}). Upper panels (left to right): Snapshot of the temperature (log), mass density (log) and Alfv\'en speed in a portion of the modelled atmosphere. Thermal instability occurs locally in a coronal loop that is in a state of thermal non-equilibrium, thereby leading to the formation of a condensation (coronal rain clump) falling towards the loop footpoint. The lower panels (same quantities in same order) show the white rectangular inset seen in the upper panels (the three  colour bars on the right of each row correspond uniquely to that particular row). Taken and adapted from Figures 2 and 3 of \citet{Antolin_2022ApJ...926L..29A}.}
	\label{Fig:Antolin2022}
\end{figure*}

Various properties of coronal rain remain poorly understood. Similar to prominences, the observed rain morphology is filamentary (also known as multi-stranded) and clumpy, whose widths (a few hundred km size) are interestingly very similar to the smallest widths detected for EUV strands in high resolution observations with Hi-C \citep{Williams_2020ApJ...892..134W}. It has been hypothesised that such sizes are set by the granular spatial scales \citep{Martinez-Sykora_2018ApJ...860..116M}. This is supported by self-consistent 2.5D rMHD numerical simulations with Bifrost (see Fig.~\ref{Fig:Antolin2022} and also Sect.~\ref{sec:rmhd}) but the details of this process remain unclear \citep{Antolin_2022ApJ...926L..29A}. Observational validation could come from multi-instrument observations with ALMA (using extended array configurations with correspondingly high angular resolution) and IRIS simultaneously observing a bright point and its indirect effects (coronal rain) on the upper atmosphere. While dense, the very small sizes of the clumps make them optically thin most of the time in spectral lines in the optical and UV spectrum such as H$\alpha$ and \ion{Mg}{ii}~k \&~h \citep{Antolin_Rouppe_2012ApJ...745..152A, Antolin_2021NatAs...5...54A}, and cannot be resolved in the present ALMA configuration. However, extended array configurations with higher resolution should allow their detection. This can be confirmed by forward modelling of Bifrost simulations with coronal rain (see Fig.~\ref{Fig:Antolin2022}), demonstrating once more the essential role of numerical simulations for developing adequate observing modes. 

Most coronal rain clumps do not occur in isolation, independently from each other. Rain clumps are produced in groups called showers \citep{Antolin_Rouppe_2012ApJ...745..152A} due to a syncing cross-field mechanism known as sympathetic cooling \citep{Fang_2013ApJ...771L..29F}. These showers have been found to successfully identify coronal loops in the `coronal veil' produced by optically thin radiation \citep{Sahin_2022, Malanushenko_2022ApJ...927....1M}. The detection and quantification of rain showers is important for coronal heating since it is a proxy for the coronal volume in TNE (and thus subject to specific spatial and temporal heating conditions). 
The detection of showers with ALMA remains a target that should be feasible at medium resolution as it is  currently offered. The fact that no rain showers have been detected with ALMA so far may be simply due to a lack of off-limb observations (with good seeing conditions) of an AR and the additional complexity of interferometric imaging off-limb. Also, low opacity of  rain showers at ALMA wavelengths might prevent their detection. The latter can and should be checked through forward modelling of numerical simulations with, e.g., Bifrost.

Coronal rain has been shown to be largely multi-thermal with probably high ionisation levels due to its relatively low lifetime. A very thin CCTR (Corona Condensation Transition Region) is expected at the boundaries of rain clumps (similar to the PCTR for prominences), with different plasma conditions ahead and behind the condensation due to a piston-like effect produced by the downward motion \citep{Antolin_2022ApJ...926L..29A}. Accordingly, the pressure restructuring is expected to be the leading cause behind the observed lower than free-fall speeds \citep{Schrijver_2001SoPh..198..325S}. An interesting mass-velocity relation has been numerically obtained but remains to be observationally validated \citep{Oliver_2014ApJ...784...21O}. The good correlation between H$\alpha$ intensity and brightness temperature measured by ALMA in prominences \citep{2022ApJ...927L..29H,2022MNRAS.513L..30L} suggests that ALMA could also serve as a proxy for prominence and coronal rain mass, and therefore shed light onto the mass-velocity relation. Indeed, the emission measure in H$\alpha$ is known to be strongly correlated to the absolute intensity in H$\alpha$ \citep{Gouttebroze_etal_1993AAS...99..513G}, and this has been used to infer the densities of rain clumps in H$\alpha$ observations with the SST \citep{Froment_2020AA...633A..11F}. 

Future ALMA observations may be able to provide detailed measurements of the morphological, thermodynamic and kinematic properties of coronal rain. Such measurements can serve as unique proxies for coronal heating conditions and improve our understanding of fundamental plasma and MHD processes such as thermal instability. As outlined above, numerical simulations would be essential for several aspects, ranging from the development of adequate observing modes and capabilities to supporting the detailed interpretation of ALMA observations.

\section{Summary and Outlook} 
\label{sec:summary}

Synthetic brightness temperature maps for millimetre wavelengths based on 3D numerical simulations of the solar atmosphere play an important role in several aspects of solar science with ALMA. First of all, such synthetic observables, for which the connection to the underlying plasma properties in the model is accurately known, have already proven their value for supporting the interpretation of ALMA observations, very much as for other chromospheric diagnostics before. 
Another important application of simulation-based observables is the development of (sub-)mm continua as a diagnostic tool for the solar atmosphere. Surprisingly though, the first ALMA observations raised questions even regarding rather fundamental properties such as the exact formation height ranges and their possible variations in time and space, locally and for different types of region. 
Also, the chromospheric temperatures in 3D simulations  \citep[see a comparison in][]{2016SSRv..200....1W} tend to be lower than the brightness temperatures observed with ALMA \citep[e.g.,][]{2017SoPh..292...88W,2017ApJ...845L..19B,2018A&A...619L...6N,2020A&A...635A..71W} although some cooler regions have been observed, too \citep{2019ApJ...877L..26L}. 
The apparent differences with respect to numerical models highlights potentially missing physical ingredients and the need for the further development of models to higher levels of realism as far as computational costs allow. 
To what extent simulations can reproduce observations and thus aid their interpretation depends ultimately on the accessible computational resources and the implementation of efficient and stable numerical schemes and the resulting affordable sizes of computational boxes, grid resolution, and considered physical processes. 
An important example in this regard is the need to account for time-dependent hydrogen ionisation as it impacts the electron densities in the chromosphere on which the millimetre continuum as mapped by ALMA sensitively depends. 
Clearly, detailed comparisons to observations, to which solar ALMA observations add a new suite of complementary diagnostics, either directly or with the powerful help of data inversion techniques, serve as crucial tests of models and as such of our current understanding of the solar chromosphere.  
In this regard it should be noted that such comparisons are limited by the realism of the  instrumental models that are used to produce artificial observations. Degradation of synthetic brightness temperature maps with a simple Gaussian would correspond to a ideal telescope with a filled aperture and ignores ALMA's sparse uv-sampling as an interferometer and also degradation effects due to Earth's atmosphere (i.e., phase errors) and the handling of noise during the imaging stage. As pointed out in Sect.~\ref{sec:degrad} such tools are under development but also require notable computational effort. 

Another valuable application of simulation-based synthetic observables is the justification of resource investments for the development of additional diagnostics and observing modes, e.g., additional ALMA receiver bands and more extended array configurations for, in principle, higher angular resolution. 
A much awaited capability of ALMA will be full polarisation measurements that would facilitate the determination of the magnetic field in the chromosphere. Detailed knowledge of the 3D magnetic field structure of the solar chromosphere (or chromospheric plasma in the corona such as coronal rain) as a function of time at high resolution would be a fundamental game-changer for understanding the dynamics, energy transport and (local) heating of this still largely elusive but yet important atmospheric layer inside ARs, QS regions, and prominences alike. 
A fundamental challenge of polarisation measurements is, among other things, the low expected degree of polarisation outside of ARs, which thus demands high sensitivity and accuracy and thus a well-developed observing mode, reliable calibration, and advanced processing/imaging routines \citep[see, e.g.,][and references therein]{2000A&AS..144..169G,Fl_etal_2015ALMA,2017A&A...601A..43L,2020IAUS..354...24W}. 
The development of these components is usually hampered by the insufficiently known ``ground truth'' against which the data should be compared to. Forward modelling of artificial polarisation measurements based on numerical 3D rMHD simulations would also here play a very important role, similar to what is attempted for continuum brightness temperatures. 

The scientific potential of ALMA observations of the Sun is promising but requires significant effort and resources to be fully unlocked due to the complexity of the instrument and the observational targets, namely the dynamic and intermittent chromosphere, both in a quiescent and active state, the atmosphere above ARs including the often dramatically evolving coronal magnetic field \citep{2020Sci...367..278F} and non-thermal electrons in solar flares (Fleishman et al. 2022 Nature, in press). 
The resulting challenges for numerically modelling of the chromosphere, prominences and the flaring corona, their appearances at the wavelengths observed by ALMA, and not at least the impact of ALMA's complex instrumental properties pose critical tests of our current understanding that will inspire future progress.

\section*{Funding}
M.S., J.C.G.G and S.W.  were supported by the SolarALMA project, which has received funding from the European Research Council (ERC) 
under the European Union's Horizon 2020 research and innovation programme (grant agreement No. 682462), 
and by the Research Council of Norway through its Centres of Excellence scheme, project number 262622. 
G.D.F. was supported in part by NSF grants AGS-2121632,  
AGS-1817277, 
and  AST-1820613 
80NSSC20K0627, 
80NSSC19K0068, 
80NSSC20K0718, 
and 80NSSC18K1128 
to New Jersey Institute of Technology.
S.G. acknowledges support from grant No. 19-16890S of the Czech Science Foundation (GA \v CR) and from project RVO:67985815 of the Astronomical Institute of the Czech Academy of Sciences. P.A. acknowledges funding from the STFC Ernest Rutherford Fellowship (No. ST/R004285/2).
H.E. was supported through the CHROMATIC project (2016.0019) funded by the Knut and Alice Wallenberg foundation.
This project has received funding from the European Research Council (ERC) under the European Union's Horizon 2020 research and innovation program (SUNMAG, grant agreement 759548). The Swedish 1-m Solar Telescope is operated on the island of La Palma by the Institute for Solar Physics of Stockholm University in the Spanish Observatorio del Roque de los Muchachos of the Instituto de Astrof\'isica de Canarias. The Institute for Solar Physics is supported by a grant for research infrastructures of national importance from the Swedish Research Council (registration number  2021-00169).

\section*{Acknowledgments}

This paper makes use of the following ALMA data: ADS/JAO. ALMA\#2011.0.00020.SV and ADS/JAO.ALMA\#2018.1.01518.S. ALMA is a partnership of ESO (representing its member states), NSF (USA) and NINS (Japan), together with NRC (Canada) and NSC and ASIAA (Taiwan), and KASI (Republic of Korea), in co-operation with the Republic of Chile. The Joint ALMA Observatory is operated by ESO, AUI/NRAO, and NAOJ. The National Radio Astronomy Observatory is a facility of the National Science Foundation operated under cooperative agreement by Associated Universities, Inc. CGGC is grateful with FAPESP (2013/24155-3), CAPES (88887.310385/2018-00) and CNPq (307722/2019-8). The NSO is operated by the Association of Universities for Research in Astronomy, Inc., under cooperative agreement with the National Science Foundation. 
Part of this work was carried out in connection with the ESO-funded Development Study for the Atacama Large Millimeter/Submillimeter Array (ALMA) \textit{High-cadence imaging of the Sun} (agreement no. 80738/18/87966/ASP). Figure~\ref{Fig:Antolin2022} was obtained through computational time granted from the Greek Research \& Technology Network (GRNET) in the 
National HPC facility ARIS. 
Part of the numerical computations were also carried out on the Cray XC50 at the Center for Computational Astrophysics, NAOJ.

\bibliographystyle{frontiersinHLTH&FPHY} 

\end{document}